\documentclass[reprint,superscriptaddress,
 amsmath,amssymb,aip,]{revtex4-1}

\usepackage{multibib}
\usepackage[colorlinks=true,%
bookmarks=false,%
linkcolor=blue,%
urlcolor=blue,%
citecolor=blue,%
breaklinks]{hyperref}

\usepackage{graphicx}
\usepackage{dsfont}
\usepackage{dcolumn}
\usepackage{bm}

\usepackage{lipsum}
\usepackage{color}
\usepackage[normalem]{ulem}
\usepackage{simplewick}
\usepackage{qcircuit}
\usepackage{braket}
\usepackage{float}
\usepackage{multirow}
\usepackage{tikz}
\usepackage[utf8]{inputenc}

\usepackage{braket}
\usepackage{subfig}

\usepackage{enumitem}

\begin{document}

\title{Structural, hydrogen bonding and dipolar properties of alkyl imidazolium-based
ionic liquids: a classical and first-principles molecular dynamics study}

\author{Ir{\' e}n{\' e} Amiehe Essomba}
\affiliation{Universit\'e de Strasbourg, CNRS, Institut de Physique et Chimie des Mat\'eriaux de Strasbourg, UMR 7504, F-67034 Strasbourg, France}
\author{Mauro Boero}
\email{mauro.boero@ipcms.unistra.fr}
\affiliation{Universit\'e de Strasbourg, CNRS, Institut de Physique et Chimie des Mat\'eriaux de Strasbourg, UMR 7504, F-67034 Strasbourg, France}
\author{Kerstin Falk}%
\affiliation{Fraunhofer IWM, Freiburg, Germany}
\author{Guido Ori}%
\email{guido.ori@ipcms.unistra.fr}
\affiliation{Universit\'e de Strasbourg, CNRS, Institut de Physique et Chimie des Mat\'eriaux de Strasbourg, UMR 7504, F-67034 Strasbourg, France}

\date{\today}

\begin{abstract}
Ionic liquids (ILs) feature a tailorable and wide range of structural, chemical and electronic properties
that make this class of materials suitable to a broad variety of forefront applications in next--generation
electronics. 
Yet, their intrinsic complexity call for special attention and experimental probes have still limitations
in unraveling the interactions occurring both in the bulk IL and at the interface with the solid 
substrates used to build the devices.
This works provides an atomistic insight into these fundamental interactions by molecular modeling to complement the information still not accessible to experiments.
In particular, we shed some light on the nature of the chemical bonding, structure, charge
distribution and dipolar properties of a series of alkyl-imidazolium-based ILs by a synergy of classical
and first-principles molecular dynamics simulations. Special emphasis is given to the crucial issue of
the hydrogen bond network formation ability depending either on the nature of the anion or on the 
length of the alkyl chain of the cation. 
The hydrogen bond strength is a fundamental indicator of the cohesive and ordering features of the ILs and,
in this respect, might be exploited to foster a different behaviour of the IL used a bulk medium or when used in electronic devices.
\end{abstract}

\maketitle

\section{Introduction}

\begin{figure*}[ht]
\centering
\includegraphics[width=0.8\linewidth]{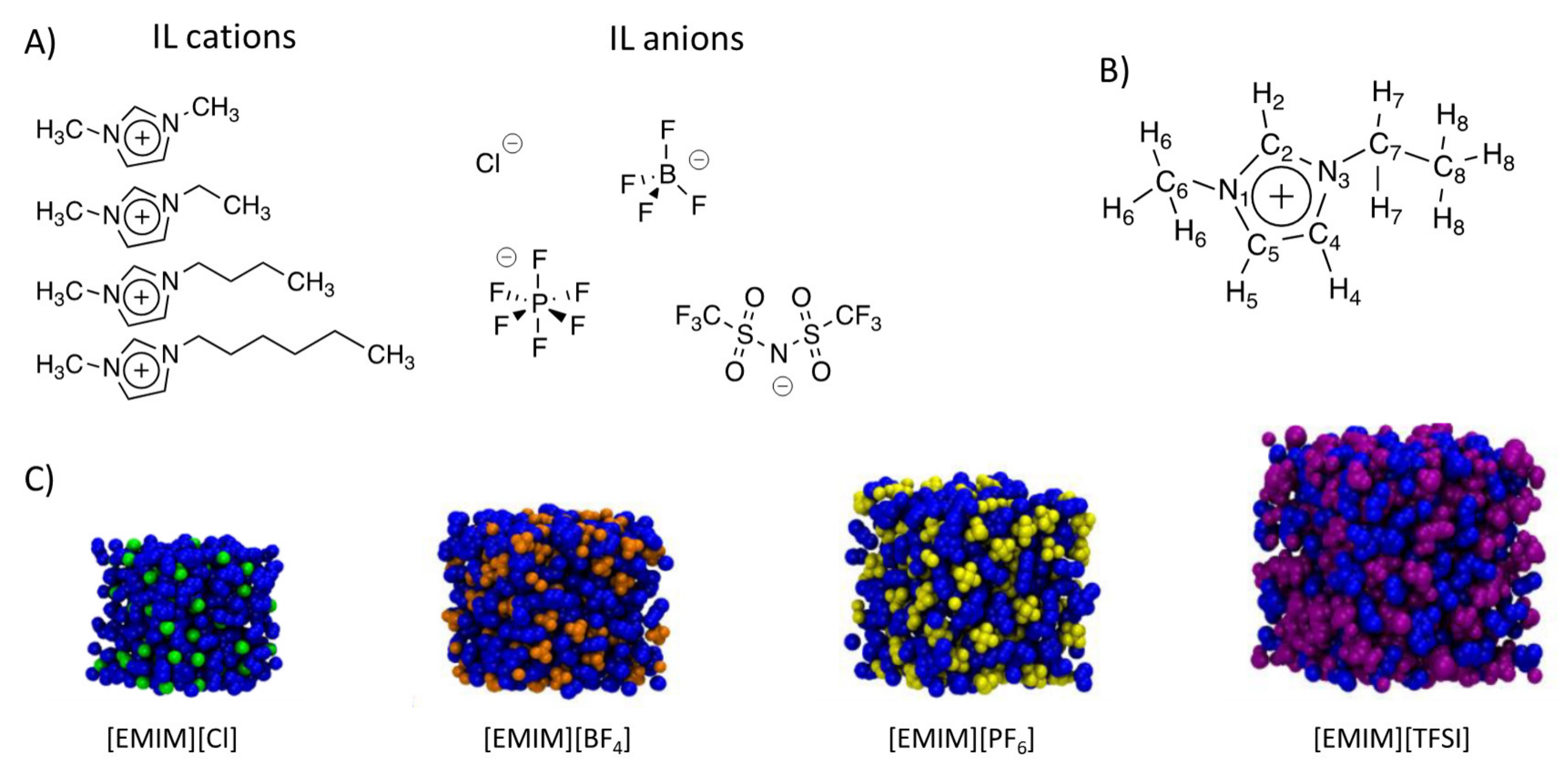} \caption{(Color online) A) Chemical structure 
of the ionic liquid cations and anions simulated in this work: as cations 1-methyl(DMIM), 1-ethyl (EMIM), 1-butyl (BMIM) and 
1-hexyl-3-methylimidazolium (HMIM) and, as anions chlorine (Cl), tetrafluoro-borate (BF$_4$), hexafluoro-phosphate (PF$_6$) and 
bis(trifluoromethylsulfonyl)imide
(TFSI). B) Example of atom indexing for 1-ethyl-3-methylimidazolium [EMIM] cation. C) Typical molecular configurations of 200 IL pairs simulated by MD.
The blue spheres are the EMIM whereas green, orange , yellow and purple correspond to Cl, BF$_4$, PF$_6$ and TFSI anions.}
\label{fig1}
\end{figure*}

Ionic liquids (ILs) bear great promise in green
chemistry, environmental science and innovative electronic devices due to their
unique properties, such as a tailorable structure, chemical variety,
and their environmentally friendly features\cite{Wang2021,Vioux2017,Welton2004,ye2012}. ILs are particularly promising for the realization of devices able to trigger interesting phenomena
such as superconductivity\cite{Morpurgo2015,Morpurgo2016,ye2012}
and quantum interference\cite{Jia2018} in 2D semiconducting (nano)materials\cite{Orgiu2016,PEI2022100159,adfm.201909736,Santato2020}.
ILs consist of cations and anions and are liquids near room temperature, they have attracted broad attention from the scientific and industrial communities because of their sizeable structural chemistry and configurable properties. 
Due to their rich interactions and microstructures, ILs may work as solvents, materials, and catalysts, rendering them broadly useful in green chemistry and novel electronic devices. Alkylimidazolium ionic liquids, especially ILs based on 1-alkyl-3-methylimidazolium cation, have been among the most studied ILs because of their remarkable properties and the simplicity of synthetic methods for producing them.
These ILs features strong coupling interactions between the electrostatic forces and hydrogen bonds with complex nano- and micro-structural ordering\cite{Welton2014,Canongia2006,Wang2022}.
In particular, the peculiar properties and functions of ILs are dominated by a unique class of hydrogen bonds.\cite{Dong2021,Dereka2021,Hunt2015} Such complexity facilitate the formation of complex nanostructures
(including ionic pairs, aggregates, or even ionic clusters) thus, the corresponding effects on the performance of ILs in typical applications are significant.\cite{Wang2022} The complex interplay between hydrogen bonds (H-bond) and dipole properties of ILs is straightly dependent on the features of the IL used. Beside their tunability in terms of chemical compositions and physicochemical properties, these features have been identified as affecting ILs performance whether used as bulk solvent phases for catalysis or as in gating media in optoelectronic transistors\cite{Wilkes2004,Nguyen2022,Matsumoto2017,Yadav2021,Park2021,Velpula}.
Atomistic modeling, especially in terms of classical molecular dynamics (MD) simulations, has been very useful in the last decade in improving our understating of ILs behaviour by complementing experimental data\cite{Canongia2004,Coasne2011,Ori2014,Ori2015}.
Several examples of MD studies oriented to the comprehension of H-bonds features in ILs are available nowadays.\cite{Morales2011,Agrawal2018,Niazi2013,Bhattacharjee2020,Andreussi2012} \textit{Ab initio} MD has also been employed to obtain further insight into the structure, H-bonds and electronic properties of a few ILs\cite{Buhl2005,DelPopolo2005,Grimme2012,Hunt2006,Cremer2010, Kirchner2015,Kirchner2010,Fayer2017}.
However these works have been focused on individual or short series of ILs.
The purpose of this work is to explore via molecular dynamics (MD) methods structural, hydrogen bonding and dipole properties of a large series of alkyl-imidazolium-based IL commonly used in optoelectronic devices.
The specific family of ILs on which we focus is constituted by imidazolium-based cations with different
lengths of the alkyl chain and the mostly experimentally exploited inorganic anions, as sketched in Figure \ref{fig1}.

Our computational study makes use of classical molecular dynamics (MD) to simulate, with the same consistent approach, a large numbers of IL systems
and to extract their structural properties and related H-bond network features. For a selected number of IL systems,
first-principles molecular dynamics (FPMD) simulations are used to enrich the information by the inclusion of
the electronic structure to refine our understanding of their interaction without relying on ad-hoc parameterized
force fields. This allow for a more precise understanding of the chemical bonding and the electrostatics
regulating the H-bond network formation and dipolar interactions in comparison with the results obtained
by classical MD.
The ability of our FPMD approach in simulating these ILs at finite temperature has already been 
assessed\cite{Kana22}, showing that this is a reliable computational protocol for the purpose of our 
present investigation.
The novelty of the study presented here stems from the relatively large number of systems considered (four cations of different alkyl chain length combined with four different anions), the 
accurate comparison of their peculiar properties and the complementary electronic properties quantified in 
terms of dipole moments, accessible by FPMD, responsible for the type and topology of H-bond network realized
by these compounds. 
The relative strength of the H-bond depends on the chemical nature and molecular structure of cations and anions
and is a fundamental indicator of the cohesive features of the ILs.
\subsection{Classical molecular
dynamics}

The ILs on which this work is focused are made of imidazolium--based cations, specifically 
1-alkyl-3-methylimidazolium cations, combined with different inorganic anions having a
general formula C$_x$MIM$^+$ -A$^-$, where x = 1, 2, 4 and 6 (also termed DMIM, EMIM, BMIM, and HMIM, respectively)
and the anion A$^-$ is chlorine Cl, or tetrafluoro-borate BF$_{4}$, or hexafluoro-phosphate PF$_{6}$
or bis(trifluoromethane)-sulfonimide TFSI.
On a general ground, a non-polarizable force fields (FF) is sufficient
to reproduce with appreciable accuracy structure and local bonding of the targeted 
ILs \cite{Andreussi2012,Boero2015,Kirchner2018,Holm2018}.
In the present work, we use the FF originally proposed and developed by Canongia Lopez and 
coworkers \cite{Canongia2004} with the parametrization of Koddermann {\it et al.} \cite{Koddermann2013}
to better account for the dynamical properties of the bulk ILs simulated. This specific FF is fitted on \textit{ab initio} calculations and allows to simulate a wide variety
of ILs chemical structures including the specific ones targeted in this work.

Periodic boundary conditions are applied to the bulk IL simulation cell and the dispersive interactions
are neglected beyond a cutoff of 15 {\AA}.
A standard Ewald summation is used in the calculation of electrostatic interaction to avoid finite 
size effects, with the Ewald parameters selected to provide an accuracy of $10^{-5}$ atomic units (au).
For each IL, we constructed two simulation cells of 20 and 200 IL pairs. This size difference will be 
clarified in the next paragraph and in the ongoing discussion. The larger systems accounting
for 200 IL pairs are sufficiently large to ensure a statistically significant average of the 
structural properties (See Table S1 in the Supplementary Material). The initial configuration has been realized as a random accommodation of IL pairs inside a cubic box by means of the packmol package \cite{packmol} and the initial volume of 
this box was selected based on the experimental density of the IL.
The bulk model of the IL was then produced following a standard melt-quench procedure heating the 
systems up to $\sim 1000$ K for 2.5 ns then by cooling these same systems to room temperature
($\sim 300$ K) in a sequence of steps in which a temperature decrease of 100 K was imposed and
on which the system was allowed to re-equilibrate for 1 ns in the canonical ensemble (NVT).
ns in the canonical ensemble (NVT).
The initial 1000 K NVT stage lasted for 2.5 ns to ensure the lose of memory of the intial 
configuration constructed as reported above.
After the NVT thermal cycle, a simulation stage in the isobaric-isothermal ensemble (NPT)
at ${\rm T}=300$ K and ${\rm P}=1$ atm was performed for 1.5 ns to get well
equilibrated models and to check the density of the simulated systems with respect to the 
experimental values. The outcome of these simulations confirmed that the theoretical density
obtained for our model ILs has small deviations from the experimental value (with the exception of DMIM-Cl ($\sim$5\%), the majority showed deviaitons $<$3\%).
These preparatory and relatively long-lasting simulations provided the actual systems simulated
for about 10 ns in the NVT ensemble. This is the production run stage in which statistics was
generated and used to fully characterize the various ILs. More precisely, the data analyzed
and presented in the following paragraphs refer to the final 5 ns in the NVT ensemble
where trajectories were sampled each 1000 steps. 
Temperature and pressure were controlled using a Nos{\' e}--Hoover thermostat \cite{Nose1,Nose2,Hoover}.
The simulations were performed with the DLPOLY package \cite{dl_poly}, and a standard Verlet
numerical integration scheme was used with a time step of 1.0 fs.

\subsection{First-principles molecular dynamics}
To account for the effects of an explicit inclusion 
of the electronic structure, we performed FPMD simulations of a few IL model systems.
To this aim, to make the system tractable within this type of approach and to mitigate
the computational workload, we used the smaller IL systems mentioned in the former paragraph.
Our FPMD is based on the density functional theory (DFT) of Kohn and Sham \cite{KS1965} (DFT)
and the dynamical simulations were performed according to the Car--Parrinello 
method\cite{CP1985} as implemented in the developer's version of the CPMD\cite{CPMDcode} code.
The valence-core interaction was described by norm-conserving numerical pseudopotential of the
Troullier-Martins\cite{TMpseudo1,TMpseudo2} type, and the exchange and correlation interactions
are expressed according to the functional proposed by Becke\cite{becke1988} and 
Lee, Yang, and Parr\cite{lee-yang-parr1988} (BLYP), respectively.
Long-range van der Waals dispersion interactions, not included in the BLYP functional, are
added according to the maximally localized Wannier functions\cite{Marzari2012} and centers (WFCs) 
approach\cite{Silvestrelli2008} boosted by a propagation scheme\cite{Boero2015}
to reduce the computational burden.
Valence electrons are treated explicitly and their wavefunctions are expanded on a plane-wave (PW)
basis set with a cut-off of 100 Ry (1360.57 eV). This value was tuned based on the convergence 
of the stress tensor (pressure) as a function of the number of PWs in the preparatory stage 
before moving to actual production run (See Figure S1 in the Supplementary Material). We remark that this relatively large cut-off is crucial
for an accurate description of the electron density around several chemical species (O, N, S, F, Cl)
composing the ILs.
Canonical NVT simulations were done by controlling the temperature via a 
Nos{\' e}--Hover\cite{Nose1,Nose2,Hoover} thermostat chain.\cite{nose1992} 
For the numerical integration of the Car--Parrinello equations of motion a fictitious electron 
mass of 400.0 au and a time step of 3.0 au (0.0726 fs) ensured good numerical control of the 
constants of motion.
The reliability of our CPMD computational protocol for the simulation of this class of systems has
already been reported elsewhere.\cite{Kana22}
The CPMD simulated ILs ([DMIM][Cl, BF$_4$, PF$_6$ and TFSI] and [EMIM][BF$_4$]) started from 
liquid phase models amounting to 20 pairs, which translates into a number of atoms between 340 and 
620, depending on the specific composition. A pre-equilibration stage within our classical MD 
scheme and following the thermal cycle explained above was the procedure adopted to generate
the initial configurations of the present set of FPMD simulations.
The stress tensor analysis done during NVT dynamical simulations at different cell sizes allowed
to assess the actual simulation cell to be used for the production and the resulting theoretical
density of the ILs turned out in fair agreement ($<$ 4\%) with the experimental value. The
system displays just a minimal residual pressure ($<\pm$ 4 GPa), (see Figure S1 of the Supplementary Material),
compatible with the DFT accuracy.\cite{DFTacc}
Trajectories were accumulated in the NVT production run at 300 K for 39 ps for [DMIM][Cl],
34 ps for [DMIM][BF$_4$], 27 ps for [DMIM][PF$_6$]), 15 ps for [DMIM][TFSI] and 20 ps for [EMIM][BF$_4$]. Representative configurations of the different IL obtained by classical MD and FPMD raw trajectory data are available at the European Center of Excellence Novel Materials Discovery (CoE-NOMAD) repository\cite{NOMAD}.

\section{Results and discussion}
\subsection{Structural properties}
\begin{figure}[htbp] 
\centering\includegraphics[width=0.7\linewidth]{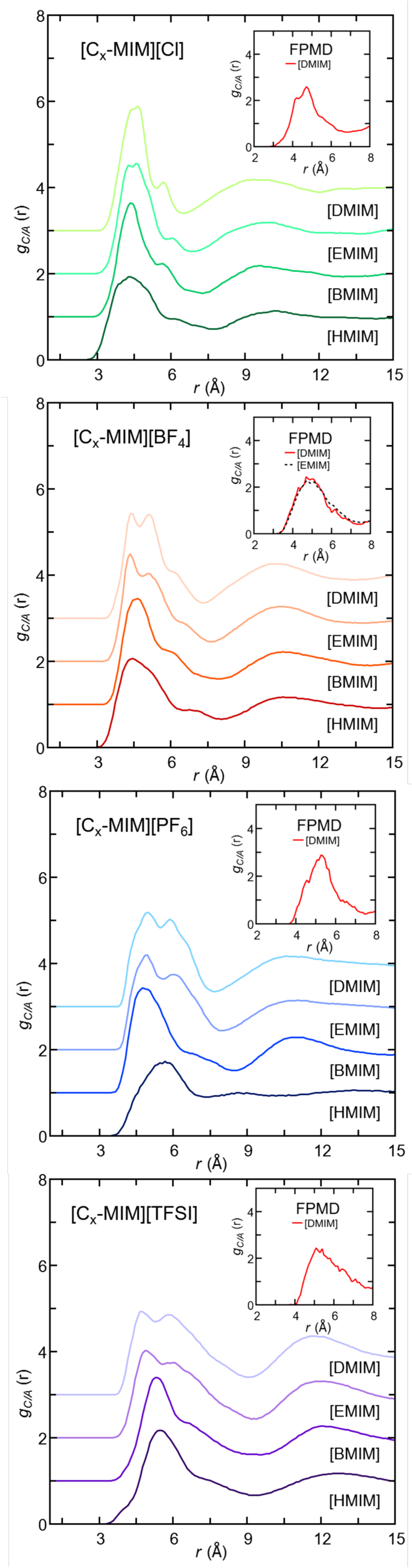}
\caption{(Color online) Cation-anion center of mass pair correlation functions (g$_{\rm {C/A}}$(r)) for the ILs simulated by classical MD at 300K (200 ILs pairs). In each plot, the insets show the g$_{\rm {C/A}}$(r) obtained by FPMD simulations (20 IL pairs).}
\label{fig2}
\end{figure}

A first issue on which we focused was the structure of the alkyl-imidazolium ILs quantified in 
terms of cation-anion pair correlation functions (PCFs) ${\rm g}_{ij}(r)$ using the standard
definition

\begin{equation}\label{gr}
    g(r) = \frac{V}{4\pi Nr^2}\sum_{i}\sum_{i\neq j}\left<
    \delta({r} - {r_{ij}})
    \right>
\end{equation}

where $N$ is the number of molecules simulated inside the cell of volume $V$ and $r_{ij}$
is the distance between the center of mass (COM) of each cation and anion constituting the 
IL. The result of these COM PCFs for the different ILs considered here are reported in 
Figure \ref{fig2}. We remind that no size-effect have been found comparing 20 and 200 ILs pairs systems with classical MD when analysing the first coordination shells around the ions (see Figure S2 in the Supplementary Material). 
These results give an insightful global picture of the structures as a function 
of both the cation alkyl chain and the type of inorganic anion. 
For the ILs in which Cl$^-$ is the anion, in the first coordination shell of the cation--anion pairs, 
the shortest alkyl chain (DMIM) shows a broad peak centered at 4.5 {\AA}, followed by a smaller peak at 
$\sim 5.8$ {\AA}. Then, by increasing the length of the alkyl chain from one to six carbon atoms, the
first peak displays an increasing broadening until the two peaks nearly merge in a unique feature for
the longer HMIM, as an effect of the larger steric volume characterizing this cation.  
For the BF$_4$ series, a similar trend is observed but overall the intensity of the peaks is reduced with
respect to the PCFs of the former ILs carrying as an anion Cl$^-$. 
The PCFs of DMIM- and EMIM-BF$_4$ show a double peak with the two maxima centered at 4.5 and 5.2 {\AA}, 
a clear indication of two preferred local coordination of the anion with respect of the orientation of the alkyl-imidazolium cation. 
In the case of BMIM, the double peak is replaced by a broad first peak is centered at 4.4 {\AA} together 
with minor second peak at 6.2{\AA}. 
Finally, the PF$_6$ and TFSI compounds show analogous trends with a further reduction in the intensity with respect to both Cl$^-$ and BF$_4$ anions. Namely, a double peak present in the case of the short alkyl chains becomes a unique broad peak upon increasing of the chain length, as seen in the former cases. FPMD simulations performed on the same ILs models (DMIM) show PCFs trends similar to those obtained by the present classical MD approach in terms of position of the first peak. A major difference is that instead of a double peak, DFT-based dynamical simulations display a single first peak but more  structured than the one obtained by classical MD.\\
An additional insightful parameter characterizing the structure of the ILs is the coordination number $n_c$. This can be obtained by integrating the PCF, $g_{ij}(r)$, as
\begin{equation}\label{}
\centering
    n_{c} = 4\pi\rho_{0}\int_{0}^{r_{c}}r^{2}g(r)dr
\end{equation}

where $\rho_{0}$ and r$_{c}$ are the atomic density of each IL pairs and the position of the first minimum 
minimum of the PCF, respectively.
The values of $n_c$ obtained from both classical MD and FPMD are reported in the table \eqref{tab_coord}.

\begin{table}\label{table1}
\caption{\label{tab_coord} \textbf{Cation-anion coordination numbers $n_c$ obtained from classical MD and FPMD 
simulations at 300K.\footnote{
Minimum of the $g_{ij}(r)$ after the first peak used as cutoff for the calculation of $n_c$ (in parenthesis the values 
used for FPMD): C$_{1,2,4,6}$MIM-Cl, 6.46(6.43) \AA, 6.86 \AA, 7.26 \AA, 7.86 \AA; C$_{1,2,4,6}$MIM-BF$_4$, 
7.25(7.32) \AA, 7.64(7.52) \AA, 8.05 \AA, 8.06 \AA; C$_{1,2,4,6}$MIM-PF$_6$, 7.68(7.46) \AA, 8.15 \AA, 8.48 \AA, 
7.26 \AA; C$_{1,2,4,6}$MIM-TFSI, 8.98(8.98) \AA, 9.17 \AA, 9.58 \AA, 9.98 \AA.}}}
\begin{ruledtabular} 
\begin{tabular}{llcccc}
&&C$_{x}$-Cl&C$_{x}$-BF$_{4}$&C$_{x}$-PF$_{6}$&C$_{x}$-TFSI\\
\hline
DMIM&MD& 
6.69(3) & 7.46(3) &   7.92(3)& 10.11(2) \\
        &FPMD              & 7.01(4) & 7.44(2) &7.12(3)  & 10.02(2) \\
EMIM& MD&6.99(2)& 7.48(3) &7.99(2) & 7.51(3)\\
         &FPMD            & -- & 7.32(3) & -- &  -- \\
BMIM&MD&6.10(3) & 6.91(2)& 7.17(4)&  7.30(2)\\
HMIM& MD&6.77(2)& 5.93(3) & 6.57(3)& 6.36(2)\\

\end{tabular}
\end{ruledtabular}
\end{table}

Using the cutoffs reported in Table \ref{table1}, the calculated total average coordination
numbers for the cation-anion in the first coordination shell fall in the range from 6 to 10.
The size and steric hindrance of the anion, which increases as Cl $<$ BF$_4$ $<$ PF$_6$ $<$ TFSI, 
mostly determine the average value of the coordination number, which turns out to be $n_c =$ 6.69, 
7.46, 7.92 and 10.11 for DMIM-Cl,-BF$_4$, PF$_6$ and TFSI, in this order.
The cation-anion coordination number for Cl-based IL does not show any particular trend 
whereas BF$_4^-$, PF$_6^-$ and TFSI show a decrease of $n_c$ upon increasing the length of the 
alkyl chain. 
A figure of merit worthy of note is the fact that the values of $n_c$ estimated from the
classical MD show a remarkable agreement with those obtained by FPMD. This provides and additional
validation of the FF used here.

\subsection{Hydrogen bonding}
\begin{figure}[ht]
\centering\includegraphics[width=0.8\linewidth]{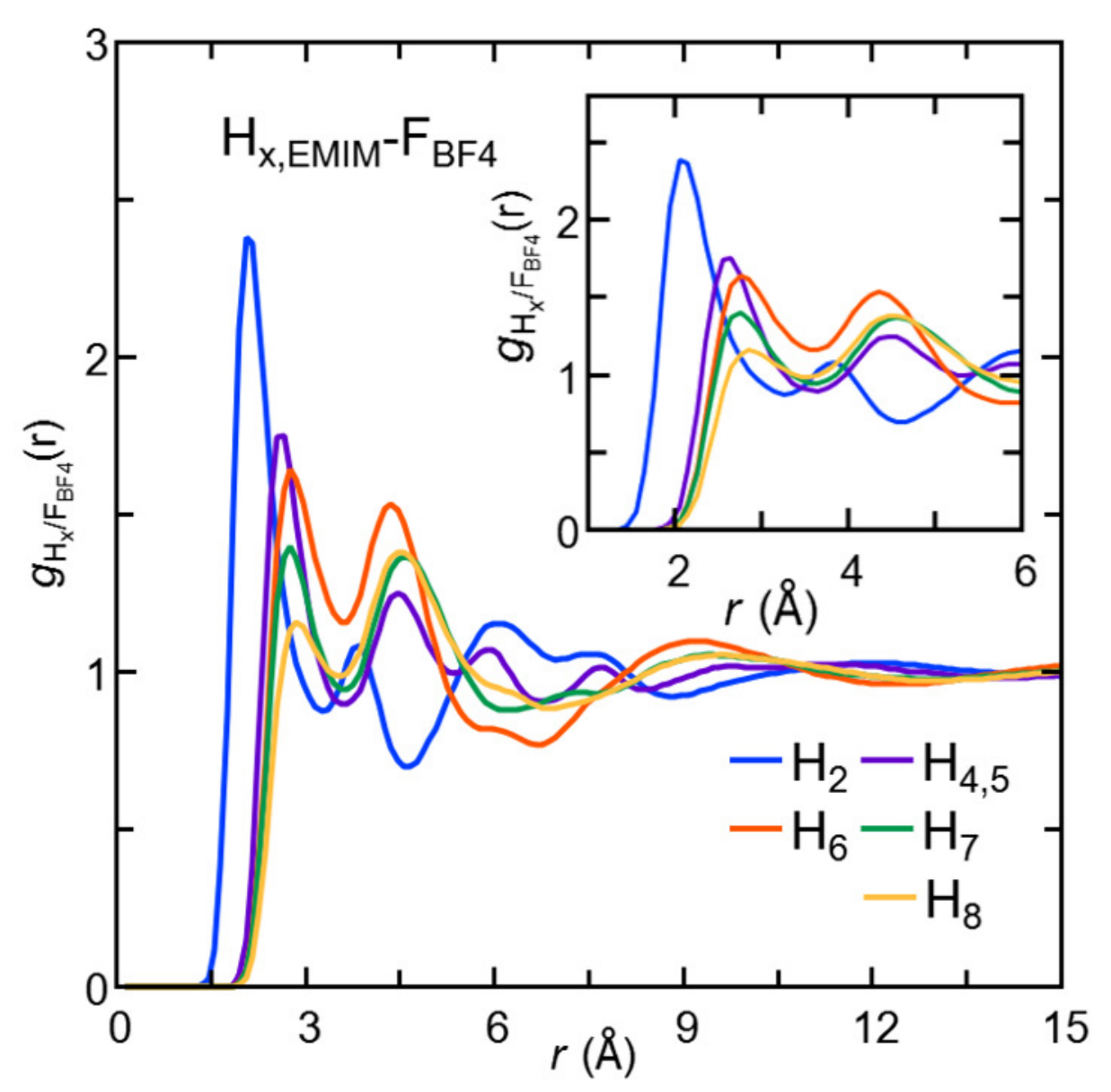}
\caption{(Color online) Pair correlation functions g$_{\rm {ij}}$(r) of the different H atoms of 
[EMIM] with the F atoms of [BF$_4$] obtained by classical MD at 300 K. The atom numbering is the standard one given in the literature and reported explicitly in Figure S2 of the Supplementary Material. Inset: a zoom-in in the 1-6 {\AA} region.}
\label{fig3}
\end{figure}

We inspected the details of the H-bond network characterizing the various ILs targeted in
this work to get a clearer picture of their nature and influence on the structure of these
systems.
A fundamental driving force determining the molecular structuring of IL is the electrostatic 
interaction occurring between IL pairs, responsible for the realization of the H-bond network
of the system and, thus, of the cohesive properties of the cation and anion moieties. 
This general observation is supported by the fact that along the evolution observed in the sampled
trajectories, the distributions of the anions (Cl$^-$, BF$_4^-$, PF$_6^-$ and TFSI) around cations 
is mainly governed by electrostatics and H-bond features, as shown by Figure \ref{fig3}, where the PCF of the 
different H atoms constituting the EMIM cation and the F atoms of the BF$_4^-$ anion are considered.
Indeed, these are the two atoms that form the inter-molecular H-bonds in this specific compound and
serve as a practical example of this type of interactions in an IL. 

\begin{figure*}[ht]
\centering\includegraphics[width=0.8\linewidth]{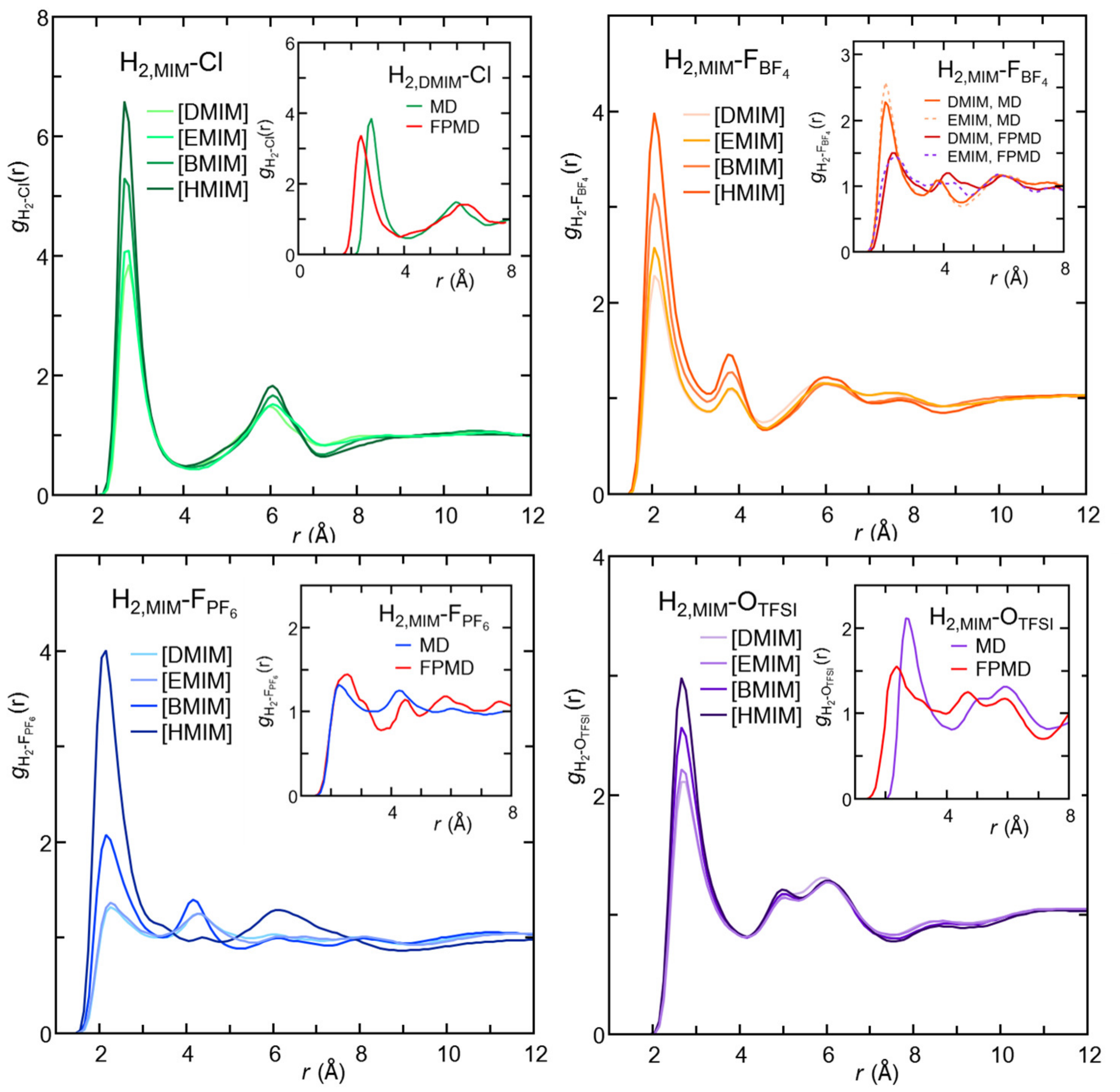}
\caption{(Color online) Pair correlation functions g$_{\rm {ij}}$(r) of the H atom (labeled as H$_2$) shared by the two 
N atoms of the IL cation and the Cl, F and O atoms atoms of the anion. The data is based on the MD simulations at 300K and is plotted for each anion as a function 
of the length of the cation alkyl chain. Inset: data obtained by FPMD simulations.}
\label{fig4}
\end{figure*}

\begin{figure*}[ht]
\centering\includegraphics[width=1\linewidth]{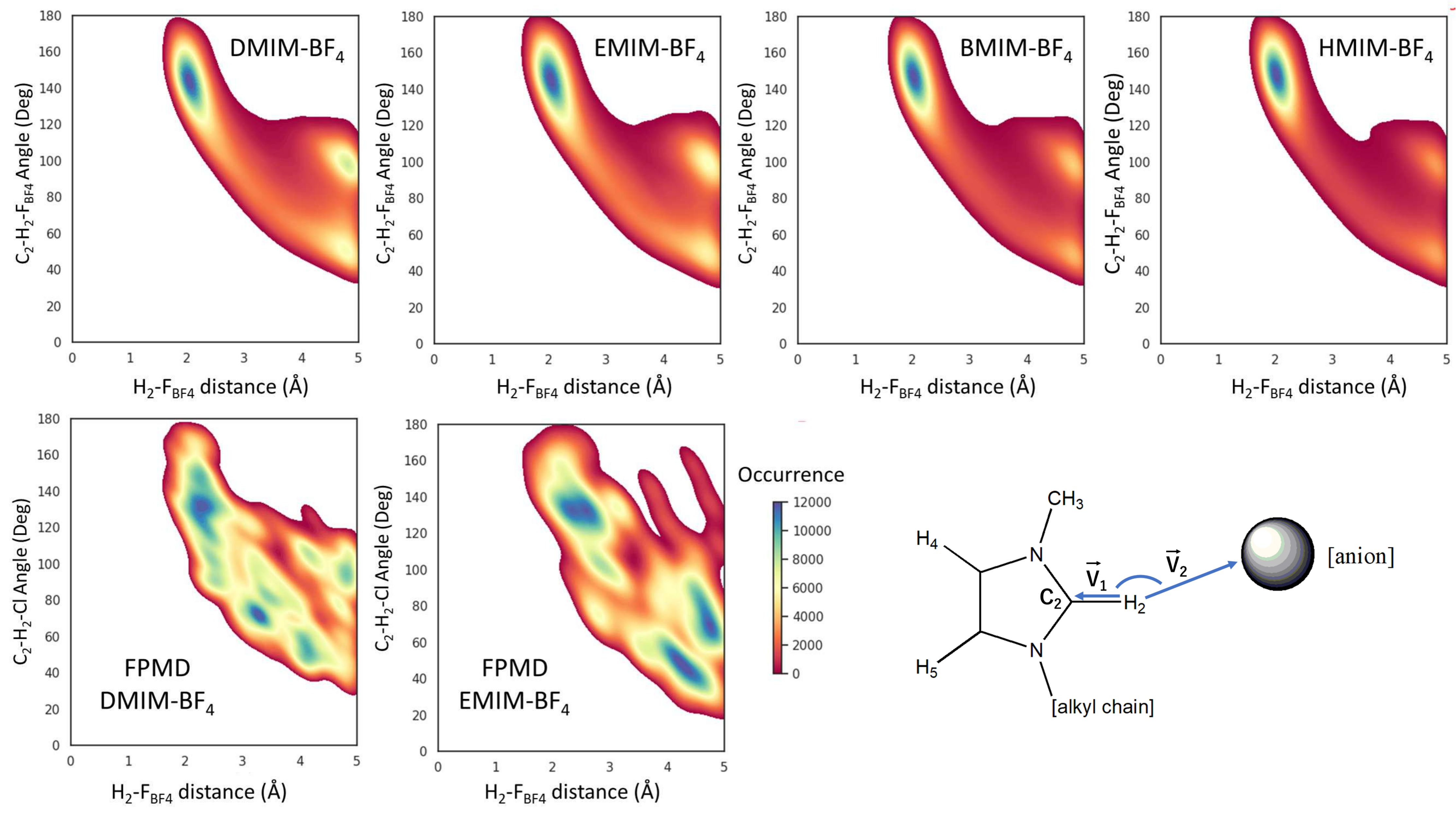} \caption{(Color online) 
Top: Combined distribution function showing the hydrogen bond
geometry between the atom H$_2$ of the IL cation and the F atoms of of the BF$_4^-$ anion as a function of the 
length of the cation alkyl chain (DMIM, EMIM, BMIM and HMIM) obtained by classical MD at 300 K. 
Bottom: Left, Combined distribution function showing the hydrogen bond obtained by FPMD; Right, Labeling of the main atoms quoted in the main text and vectors defining the angle for which distribution density plots are shown in the manuscript.
}
\label{fig5}
\end{figure*}

By examining the PCF of Figure \ref{fig3}, we can see that the sharp peak realized by H$_2$, which is
the H atom shared by the two N atoms of the imidazolium ring, in the g$_{H_2F}$(r) distribution is located
at 2.05 \AA. This is a clear indication of a strong preference of F atoms for binding to this site. 
Such a short distance is also a fingerprint of a tight H-bond.\cite{HbondILs,HbondILs2}
The other H atoms of the cation are characterized by corresponding PCFs peaked at larger distances, namely
2.5--2.7 {\AA} and reduced intensities with respect to g$_{H_2F}$(r). 
These are still active H-bonds, concurring to the overall structure of the IL \cite{HbondILs2}, but
being somehow weaker, they can be broken and reformed in a dynamical way as observed in the case of
[EMIM][TFSI].\cite{Kana22}
Analogous results for the other ILs targeted in this study are reported in Figure S3 of the 
Supplementary Material. However, with respect to [EMIM][BF$_4$] some differences can be pointed out. In particular, [EMIM][Cl] PCFs shows that together with H$_2$ atom also the other H atoms within the imidazolium ring as well as the H atoms of the first C atoms connected to the ring (C$_6$ and C$_7$) show clear sharp peaks. These H-bond interactions are found at larger distance with respect to to the case of F in [EMIM][BF$_47$], being at 2.65 {\AA} (H$_2$,H$_{4,5}$) and 2.85 {\AA} (H$_6$,H$_7$) for respectively. In the case of [EMIM][PF$_6$], the interaction between F atoms and H$_2$ atoms is found to be again the strongest with respect to the other H atoms of the cation (2.25 {\AA} versus 2.75--2.95 {\AA}), however the difference in intensity of the different H is much less pronounced with respect to [EMIM][BF$_4$]. In the case of [EMIM][TFSI], the strongest H-bond interaction is found to be the one between H atoms of the cation and the O atoms of the sulphonyl groups of TFSI. In this case, the H$_2$-O PCF first peak is found to be the most intense however all the H$_x$-O show a H-bond interaction at a similar distance (2.66--2.7{\AA}).\\
The effect of the alkyl chain length is summarized in Figure \ref{fig4}. A feature that emerges is a
structuring effect of the H-bond promoted by H$_2$ with the increase of the number of C atoms composing
the alkyl chain, from DMIM to the longer HMIM. 
This alkyl chain length effect is present in all the four ILs differing in the anion moiety. The main 
difference is a decrease in the intensity of the first peak according to the trend Cl$ >$ BF$_4 >$ 
PF$_6 >$ TFSI. 
The comparison with the FPMD results for the DMIM and EMIM systems, confirms the trend evidenced by classical MD
but reveals also the presence of slightly shorter H$_2$-Cl and H$_2$-O distances in the cases of DMIM-Cl and DMIM-TFSI, respectively. \\
In Figure \ref{fig5} we show the combined distribution function (CDF), in which the horizontal axes represent the distance 
of the hydrogen atom H$_2$ of the cation form the F site of the BF$_4^-$ anion participating to the formation of H-bond.
The vertical axes of each panel, instead, show the angle defined by the vector oriented from H$_2$ to the C$_2$ atom 
of the cation and the vector connecting H$_2$ and F as sketched in Figure \ref{fig5}.
In the angle analysis presented in Figure \ref{fig5} value of 180$^\circ$ indicates that the atoms C$_2$, 
H$_2$ and F are aligned, thus the H-bond results perfectly linear. 
An intense maximum arises in the region around 2 {\AA} for a range of value between 125$^\circ$ and 180$^\circ$ 
for the IL pair on which we focus. This originates from the H-bond formed by the H$_2$ atom of the cation and 
and the acceptor F atom of the BF$_4$. By increasing the length of the alkyl chain, a narrowing of the dispersion 
of H-bonding appears in the region around 2 {\AA} and angles of 125$^\circ$-180$^\circ$. 
This is in perfect agreement with the features displayed by the PCFs in Figures \ref{fig3} and \ref{fig4}.  
The lower panels of Figure \ref{fig5} shows the CDF data obtained from FPMD data for DMIM-BF$_4$ and EMIM-BF$_4$ systems. The FPMD data show a position of the strong H-bond at values similar to those computed from the classical MD simulations,
but the distributions are less localized and affected by a larger broadening. 
This can be ascribed both to the oscillations of the electronic structure and to the reduced statistics
of FPMD simulations in terms of system size and simulation time scales.
\begin{figure*}[ht]
\centering\includegraphics[width=1\linewidth]{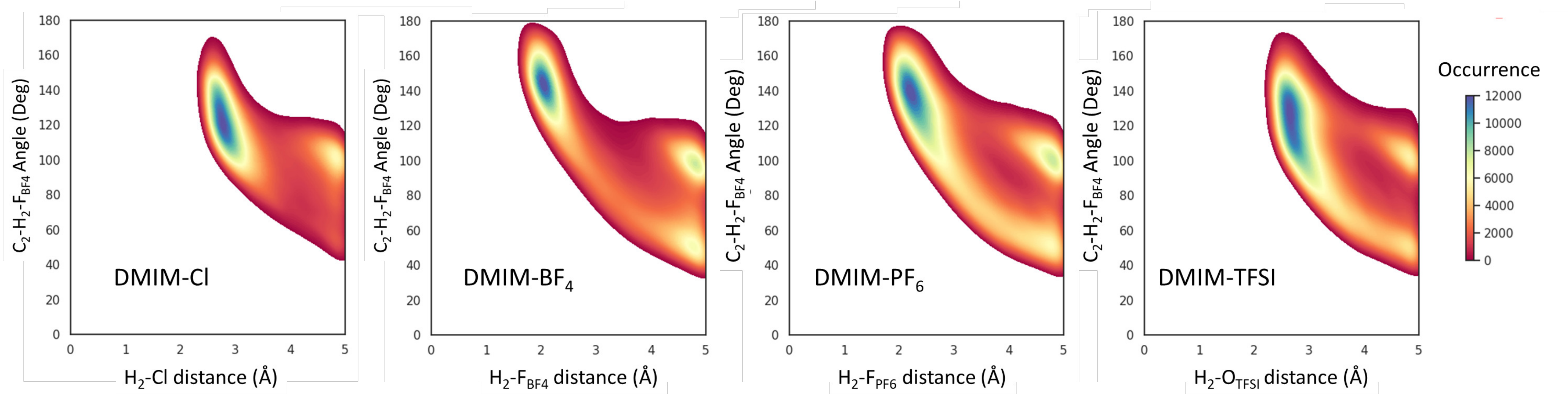} 
\caption{(Color online) Combined distribution function showing the hydrogen bond
geometry between the atom H$_2$ of the DMIM and the Cl, F and O atoms of the anions obtained by classical MD at 300 K.}
\label{fig6}
\end{figure*}
\\
Figure \ref{fig6} reports analogous two-dimensional maps of the CDFs in the case of different anions at fixed (DMIM) cation. Note that for BF$_4^-$ and PF$_6^-$ the accpetor anion atom is F whereas for TFSI is the sulphonyl O atom. In this case, corresponding to the shortest alkyl chain considered here, the differences in these distributions
can be clearly ascribed to the nature of the anion. We remark generally well localized CDF peaks, with the strongest
localization corresponding to the BF$_4^-$ for a distance of about 2 \AA and an angle of 150$^\circ$.  
Broader distribution characterize, instead, the cases of Cl$^-$ and TFSI in terms of angles which span the intervals
100$^\circ$-160$^\circ$ and 85$^\circ$-160$^\circ$ respectively. This is not entirely surprising. Indeed, the small
Cl$^-$ anion is more mobile and can find stable positions in a slightly wider angular range than BF$_4^-$.
On the other hand, both the larger PF$_6^-$ and the flexible TFSI molecule, the latter having also different
conformers\cite{Kana22}, require more space to be accomodated beside DMIM cations and this accounts for a
more dispersed CDF distribution.

\subsection{Dipole moments}
\begin{figure}[ht]
\centering\includegraphics[width=0.9\linewidth]{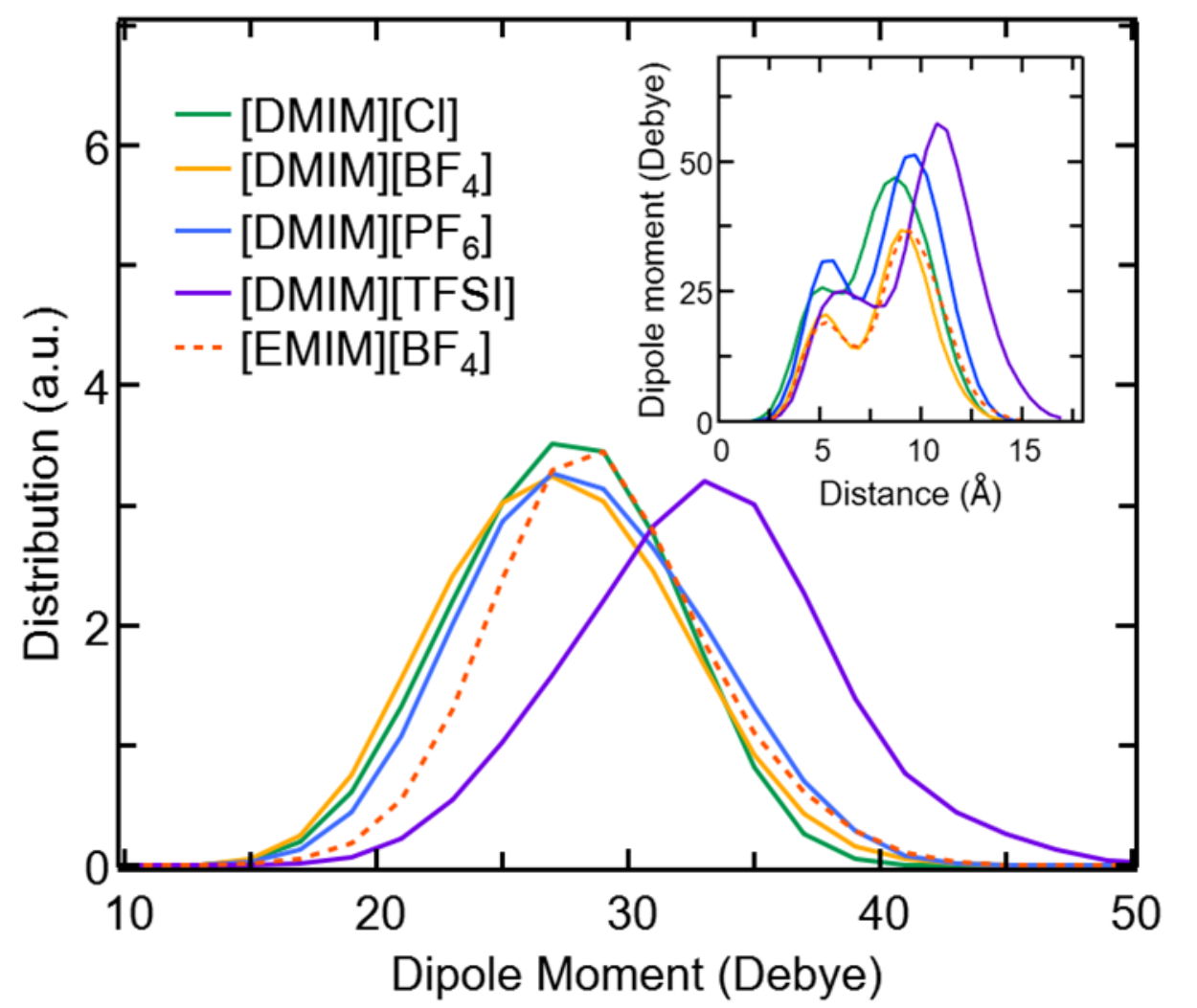} \caption{(Color online) Averaged dipole moment distributions of the IL pairs inside the different ionic liquid. Inset: Averaged dipole moment distribution plotted as a function of the distance of two counter ions.}
\label{fig7}
\end{figure}

In the case of FPMD simulations, the availability of the electronic structure allows for a straightforward
calculation of the dipole moment. To this aim, we make use of the maximally localized Wannier functions
and centers (WFCs) \cite{Marzari2012}, consisting in a unitary transformation of the Kohn–Sham orbitals.
The WFCs provide a useful tool to analyze the chemical bonds and a shorthand visualization of the
electronic structure. An example of WFCs representation for the different cations and anions of the ILs 
targeted here is reported in Figure S4 of the Supplementary Material.
The WFCs allow also for an easy calculation of the local dipole moment ${\bf \mu}$ of each cation-anion pair
constituting the IL in terms of a practical point-like charges sum, namely

\begin{equation}
\centering
    {\vec \mu} = \sum^{N}_{i=1}q_{i}{\bf R}_{i} - \sum^{n_{WFC}}_{s=1}f_s{\bf r}_{s} 
\end{equation}

where $N$ is the number of atoms in a IL ion, $q_i$ is the valence (in the case of pseudopotentials approaches)
atomic charge of atom $i$, ${\bf R}_i$ are the coordinates of atom $i$, $n_{WFC}$ is number of WFCs and $f_s$ is the
occupation of the $s^{\rm th}$ Wannier orbital, namely $f_s=2$ in a spin-restricted calculation and 
$f_s=1$ in the spin-unrestricted case, and ${\bf r}_s$ are the cartesian coordinates of each WFC belonging 
to an IL cation or anion. 
The analysis of the value of the dipole moment $\mu=|{\vec \mu}|$ was performed on the basis of MLWFs 
computed over YYY uncorrelated configurations for each IL system sampled along the dynamical trajectories 
obtained at 300 K for [DMIM]-[Cl], -[BF$_4$], -[PF$_6$], -[TFSI] and [EMIM][BF$_4^-$]. 
We recall that the only gauge invariant dipole having a physical meaning in charge neutrality condition of 
the simulation cell is the one between the cation and the anion.\cite{Kirchner2011}
In our analysis, we computed the dipole moments between each cation-anion pair present in our simulation
cells and the statistics is extracted based on both the number of pairs and the different configurations
sampled along the trajectories. The result is shown in Figure \ref{fig7}.

The distribution of the dipole moment show an average value of 27.0 D with a full width at half the maximum (FWHM)
of about 10 D for both [DMIM]-[BF$_4^-$] and [DMIM]-[PF$_6^-$]. In the case of [DMIM]-[Cl$^-$], while the FWHM does
not undergo any change, the distribution shifts toward slightly higher values with a peak centerd at 28.0 D.
The most remarkable displacement of the distribution was found for the case of [DMIM]-[TFSI], where the average dipole
moment takes the value of 33.0 D, consistently with the more bulky anion structure.
A general picture that can be extracted is the relative stability and insensitivity of the local dipole moment 
to the type of anion (Cl, BF$_4$, PF$_6$), provided that such an anion has not a complex molecular
structure, for the same IL cation (DMIM). An analogous minimal effect has been found for different alkyl chains
in the cases of DMIM and EMIM when BF$_4$ is the cation. 
The larger dipole moment found in the case of [DMIM]-[TFSI], as opposed to the other cases presented in Figure \ref{fig7}
is mainly due to the more complex molecular structure of the [TFSI] with respect to the other smaller and roughly 
spherically isotropic anions (Cl, BF$_4$, PF$_6$). The bigger TFSI molecule has a larger steric encumbrance
and is accommodated at larger distances from the cation, as shown in the analysis of the PCFs of Figure \ref{fig1}.
As a consequence, the relative distance between the center of mass of the positive and negative charges increases,
thus accounting for an increased dipole moment, in line with the results obtained for [EMIM]-[TFSI].\cite{Kana22}

As expected, the presence of an extended H--bond network induces a significant enhancement of the dipole moment
of a corresponding dimer in the gas phase
(14.3 D, 13.6 D, 13.9 D, 14.1 D, 17.1 D, for [DMIM]–[Cl], [DMIM]–[BF4], [DMIM]–[PF6], [DMIM]–[TFSI] and [EMIM]–[BF4], 
respectively). 
This enhancement, typical of any polar liquid, is again an indicator of the presence of H-bonds and a 
fingerprint of the crucial electrostatic interactions between cation and anion\cite{Kana22}.  

In Figure \ref{fig7}, we report in the inset also the values of $\mu$ as a function of the interionic distance 
of the cation-anion pair extending up to the second coordination shell (0-15 {\AA}). 
A double-peak feature characterizes the IL in which the anions are Cl$^-$, BF$_4^-$ and PF$_6^-$, with the two
maxima centered at $\sim 5$ and $\sim 9$ {\AA}.
Since the dipole moment of an ionic liquid is experimentally difficult to measure, this data can provide a useful 
quantitative analysis to complement experimental probes.

\section{Conclusions}
We have inspected a series of alkyl-imidazolium-based ILs, targeted in forefront electronic applications, with the scope
of unraveling their atomistic structure, specific interactions and microscopic features. By resorting to a combination of classical and first-principles dynamical
simulations, we have shown how the composition of the ILs influences the structure, the nature and topology of the
H-bond network and, in turn, the electronic properties, quantified in terms of dipole moment distribution of the 
different ILs. Our study makes use of an identical simulation protocol, based on both classical model potentials
and DFT--based first-principles molecular dynamics, of a series of ILs on which forefront research is focused.
This comparative study, done within an identical and consistent paradigm is prone to give a useful guidelines in the selection of the best suited IL systems according
to the specific application targeted where the ability of H-bond may play a significant role.

\section*{Supplementary Material}
The Supplementary Material include data and supporting figures quoted in the text.

\section*{Data availability}
Representative configurations of the different IL obtained by classical MD and FPMD raw trajectory data are available at the European Center of Excellence Novel Materials Discovery (CoE-NOMAD) repository.\cite{NOMAD}

\begin{acknowledgements}
We gratefully acknowledge computational resources at the HPC Centre of the University of Strasbourg
funded by the Equipex Equip@Meso project (Programme Investissements d’Avenir) and CPER Alsacalcul/Big Data, and at Grand Equipement National de Calcul Intensif (GENCI) under allocation DARI A0100906092 and A0120906092.
G.O. and K.F. acknowledge the European Campus Eucor initiative for Seed Money funding (Project MEDIA).
Funding: I.E.A. is funded
by the Idex Contrats Doctoraux – Programme Doctoral International (N. ESRH0908292D)
of the University of Strasbourg.
We thank Laurent Douce (University of Strasbourg) and Carlo Massobrio (University of Strasbourg), Emanuele Orgiu (INRS, Canada) and Jocelyne Levillain (University of Caen) for fruitful discussions.
\end{acknowledgements}


\end{document}